# WHO SETS THE RANGE? FUNDING MECHANICS AND 4H CONTEXT IN CRYPTO MARKETS

**Prof. Habib Badawi[1]**, **Dr. Mohamed Hani[2]**, **Dr. Taufikin Taufikin[3]**

[1]Lebanese University, Beirut, Lebanon (habib.badawi@ul.edu.lb)
[2]University of Akli Mohand Oulhaj-Bouira, Algeria (m.hani@univ-bouira.dz)
[3]Universitas Islam Negeri Sunan Kudus, Indonesia (taufikin@iainkudus.ac.id)

## ABSTRACT

Financial markets are often described as chaotic, with price movements appearing erratic and driven by sudden sentiment shifts. This interpretation obscures a fundamental reality: ranges are rarely accidental. They are structured outcomes shaped by underlying market context and capital conditions that govern price behavior.

This paper advances the argument that the four-hour (4H) timeframe offers a critical analytical lens for understanding market structure. Positioned between intraday fluctuations and higher-timeframe macro trends, the 4H context captures the equilibrium zone where institutional positioning, leveraged exposure, and liquidity management converge. Within this temporal window, markets reveal their true architecture—one governed less by headlines and more by the interaction between structural context and funding dynamics.

Funding mechanisms, particularly in derivative-driven markets, operate as a subtle but powerful disciplinary force. They regulate trader behavior, incentivize positioning, and impose economic costs that shape directional commitment. When funding aligns with the prevailing 4H market context, price expansion becomes possible. When it diverges, compression and range-bound behavior emerge. Funding does not merely accompany price action—it actively constrains and channels it, functioning as a governor on market movement.

By reframing market ranges as products of contextual alignment and funding pressure, this paper challenges the narrative that attributes consolidation phases to indecision or randomness. Instead, ranges represent periods of controlled balance—moments when capital, leverage, and structural context reach temporary equilibrium. These phases are not transitional voids between trends but meaningful states reflecting strategic positioning by informed market participants.

This analysis calls for a more disciplined reading of market behavior—one that privileges structure over spectacle and mechanics over mythology. Understanding how 4H context and funding operate as market governors is essential for interpreting price action as a rational, power-mediated process rather than an unpredictable phenomenon.

# 1. INTRODUCTION: THE ARCHITECTURE OF GOVERNED MARKETS

This paper argues that market ranges—particularly in highly leveraged and derivative-driven environments—are neither incidental nor disorderly. They are structured outcomes produced by the interaction of temporal context and capital mechanics. When examined through the 4H timeframe, price behavior reveals a patterned logic in which market equilibrium is actively maintained rather than passively observed. What appears as stagnation is the visible expression of hidden constraints.

The 4H context functions as a structural midpoint where short-term volatility and long-term directional narratives intersect. At this level, institutions continuously recalibrate liquidity deployment, risk management, and leveraged exposure. Price is permitted to fluctuate, but only within boundaries that reflect acceptable risk and funding cost. These boundaries define the range. Far from arbitrary, they represent the market's way of enforcing discipline against competing incentives.

The selection of the 4H timeframe emerges from practical realities of how modern cryptocurrency markets operate. This temporal window represents the junction where intraday liquidity games meet daily, and weekly regime shifts and where leverage resets are felt without being washed out by macro cycles. Within this frame, boundary formation occurs through the convergence of acceptable risk parameters and funding cost ceilings.

The upper bound of any range often coincides with liquidity shelves—price levels where significant orders cluster on exchange order books—and short-gamma zones, areas where market makers have positioned inventory and where sudden demand encounters resistance. The lower bound aligns with collateral resilience thresholds and long-gamma cushions, representing the floor beneath which forced liquidations cascade and destabilize the structure.

This equilibrium function means the 4H chart effectively encodes the corridor that institutions tolerate while financing positions, managing liquidation bands, and resetting risk exposure. It is tight enough to constrain reckless speculation yet broad enough to absorb normal trading activity. What appears as governed drift—price circulating within seemingly stable bounds—is determined by liquidity availability, margin elasticity, and the cumulative cost of keeping imbalanced exposure alive. Markets move deliberately within boundaries negotiated by capital flows and structural constraints.

# 2. KEY DEFINITIONS AND CONCEPTUAL FRAMEWORK

Before proceeding, several technical terms require clarification. These foundational concepts establish the analytical vocabulary necessary for understanding how 4H context and funding dynamics govern cryptocurrency market behavior. The technical terminology presented here represents more than mere jargon; these concepts constitute the essential building blocks for understanding market microstructure in perpetual futures markets. Table I systematically organizes these core definitions according to their market function, enabling readers to grasp how each component contributes to the broader framework of market governance.

### Table I. Core Technical Definitions and Market Functions

| Term | Definition | Market Function |
|---|---|---|
| 4H Timeframe | Four-hour candlestick charts that aggregate price data over four-hour periods | Filters intraday noise while remaining responsive to institutional positioning changes |
| Funding Rate | Periodic payment exchanged between long and short position holders in perpetual futures to maintain parity with spot prices | Positive funding: longs pay shorts; Negative funding: shorts pay longs |
| Open Interest | Total number of outstanding derivative contracts (futures or options) not yet settled | Rising OI with stable prices suggests accumulation; falling OI suggests deleveraging |
| Liquidation Cascade | Chain reaction where forced closures of leveraged positions trigger price movements forcing additional liquidations | Creates rapid and amplified price changes through feedback loops |
| Perpetual Futures | Derivative contracts without expiry dates tracking spot prices through funding mechanisms | Maintains price parity through continuous funding rate adjustments |
| Gamma Exposure | Rate of change in delta (price sensitivity) in options markets | Short gamma zones create resistance; long gamma zones provide stability |
| Market Structure | Arrangement of price levels, volume profiles, and liquidity defining market operations | Includes support/resistance zones, accumulation/distribution patterns, trending vs. ranging behavior |
| Liquidity Shelf | Price levels where significant orders cluster on exchange order books | Creates natural support/resistance through concentrated liquidity |
| Basis Spread | Price difference between derivative contracts and spot prices | Indicates arbitrage opportunities and funding tightness |
| Collateral Elasticity | Capacity of the system to accommodate position sizes given available collateral | Determines carrying capacity of positions and range width |

*Note*: These definitions establish the conceptual vocabulary for understanding 4H context and funding dynamics in cryptocurrency markets. Each term represents a measurable component of market architecture amenable to empirical analysis.

The funding rate merits particular emphasis as it operates not simply as a fee structure but as a sophisticated parity mechanism that continuously reconciles derivative and spot prices through economic incentives. When perpetual contracts trade above spot prices, positive funding rates extract payments from long positions and transfer them to shorts, creating economic incentive for arbitrageurs to sell perpetuals and buy spot, thereby restoring equilibrium. This mechanism transforms abstract positioning imbalances into concrete costs that accumulate over time, imposing discipline on speculative excess long before price collapses become necessary.

Similarly, the 4H timeframe filters intraday noise while preserving sensitivity to institutional positioning shifts—a temporal resolution that proves optimal for capturing the equilibrium zone where strategic capital deployment becomes observable. Unlike one-minute charts that reflect algorithmic noise or daily charts that respond sluggishly to structural changes, the 4H window reveals deliberate institutional behavior while maintaining sufficient responsiveness to regime transitions.

# 3. THEORETICAL FRAMEWORK

## 3.1 Market Microstructure and Price Formation

This study builds upon foundational market microstructure theory, which examines how trading mechanisms, information asymmetry, and institutional behavior influence price discovery and formation (O'Hara, 1995). The analytical focus on the 4-hour timeframe aligns with Kyle's (1985) sequential trading model, where informed traders strategically time their participation to minimize price impact while maximizing returns. Kyle's framework establishes that price reflects the interaction between informed traders, noise traders, and market makers—a dynamic particularly visible at intermediate temporal resolutions where strategic positioning becomes observable.

The paper's emphasis on temporal scale selection echoes the fractal market hypothesis proposed by Peters (1994), which argues that different investment horizons create distinct market structures. The 4-hour timeframe represents a critical junction where short-term speculative flows meet institutional positioning cycles, creating what Müller et al. (1997) identified as a "heterogeneous information arrival" zone where multiple participant classes interact most intensively.

## 3.2 Liquidity Dynamics and Funding Constraints

The paper's treatment of funding mechanisms as disciplinary forces draws directly from Brunnermeier and Pedersen's (2009) seminal work on market liquidity and funding liquidity spirals. Their model demonstrates that when funding becomes constrained, traders reduce positions regardless of fundamental value, creating feedback loops between market liquidity (ease of trading assets) and funding liquidity (ease of obtaining capital). This framework explains why elevated funding rates compress price ranges even when directional narratives remain compelling—the cost of capital imposes economic discipline that overrides conviction.

Gromb and Vayanos (2002) extend this analysis by demonstrating how arbitrageurs' capital constraints affect price efficiency and create persistent mispricings. In cryptocurrency markets, where leverage ratios regularly exceed 100x and funding rates reset every eight hours, these constraints operate with intensity, making Gromb and Vayanos's limits-of-arbitrage framework especially applicable.

## 3.3 Derivatives and Leverage Dynamics

The paper's analysis of perpetual futures and funding rates connects to the broader literature on derivatives' impact on underlying markets. Chan and Fong (2006) demonstrate that derivative trading can enhance or impede price discovery depending on leverage usage patterns and the balance between informed and uninformed participation. Perpetual futures, unique to cryptocurrency markets, introduce continuous rebalancing mechanisms through funding rates that have no direct parallel in traditional finance.

Ni, Pearson, and Poteshman (2005) show how options market positioning creates asymmetric hedging pressure that influences spot prices—a phenomenon this paper extends to the "gamma exposure" zones identified in the 4-hour context. Market makers' dynamic hedging requirements create what Garleanu, Pedersen, and Poteshman (2009) term "demand-based option pricing," where positioning imbalances affect derivatives prices independent of fundamental factors.

## 3.4 Game-Theoretic Foundations of Strategic Positioning

The distributed governance mechanism described in the paper aligns with game-theoretic models of strategic interaction in financial markets. Carlin, Lobo, and Viswanathan (2007) model how traders' episodic attention and information processing constraints create predictable patterns in order flow and liquidity provision. The 4-hour timeframe may represent an equilibrium observation window where the costs of continuous monitoring are balanced against the benefits of responding to structural shifts.

Vayanos and Wang (2013) provide a comprehensive framework for understanding how search frictions, bargaining power, and capital constraints interact to determine asset prices. Their model explains why markets exhibit "preferred habitats" at price levels—the liquidity shelves and gamma zones identified in this paper—as emerging from the distribution of trader preferences, capital endowments, and information sets.

## 3.5 Regime-Dependent Market Behavior

The paper's distinction between accumulation, distribution, and trending phases connects to regime-switching models in financial econometrics. Hamilton (1989) pioneered "Markov-switching models"[1] that allow market parameters to shift between discrete states, while Ang and

---

[1] A **Markov-switching model** is a regime-switching technique that allows for the parameters of a time series to change across discrete, unobserved states, such as "low" versus "high" volatility regimes. In financial markets, this framework

Bekaert (2002) demonstrate how regime identification improves risk measurement and forecasting. The paper's framework suggests that funding alignment serves as a regime indicator, signaling when markets can transition from range-bound to trending behavior.

Pagan and Sossounov (2003) establish methodologies for dating market cycles and identifying turning points—techniques applicable to validating the paper's hypothesis that funding moderation precedes genuine breakouts while sustained funding pressure extends consolidation phases.

## 3.6 Cryptocurrency-Specific Market Structure

Recent scholarship specifically examining cryptocurrency markets provides additional grounding. Makarov and Schoar (2020) document systematic price fragmentation and arbitrage failures across cryptocurrency exchanges, demonstrating that even basic efficiency mechanisms encounter friction due to capital constraints and regulatory segmentation. Their findings support the paper's emphasis on collateral cycles and funding liquidity as determinants of market behavior.

Griffin and Shams (2020) identify price manipulation patterns in cryptocurrency markets, showing how concentrated actors can influence prices through strategic trading, particularly during low-liquidity periods. This manipulation capability makes range boundaries particularly consequential—they represent not just technical levels but zones where strategic actors concentrate efforts to defend positions or trigger cascades.

Gandal et al. (2018) document price manipulation through wash trading and coordinated buying on Mt. Gox, establishing that cryptocurrency markets remain vulnerable to artificial price inflation. The paper's framework of "power-policed boundaries" acknowledges this reality while arguing that funding mechanisms provide countervailing discipline against sustained manipulation.

## 3.7 Information Processing and Temporal Aggregation

The paper's privileging of the 4-hour timeframe relates to literature on optimal information aggregation across time scales. Andersen and Bollerslev (1997) demonstrate that intraday volatility patterns exhibit strong periodicity and heteroskedasticity, making raw high-frequency data unsuitable for structural analysis without appropriate filtering. The 4-hour aggregation represents a practical compromise between signal extraction and responsiveness.

Hasbrouck (1991) establishes that different trading venues and time horizons contribute differentially to price discovery, with some time scales dominated by noise while others capture genuine information revelation. The paper's framework suggests the 4-hour scale optimally balances these considerations for cryptocurrency markets where continuous trading, global participation, and high leverage create unique information dynamics.

---

is used to identify non-linear transitions between market phases, such as moving from a stable range-bound environment to a trending state.

## 3.8 Synthesis: Toward an Integrated Framework

These theoretical foundations converge to support the paper's central argument: cryptocurrency market ranges represent governed equilibria emerging from the interaction of funding constraints, liquidity positioning, and temporal context rather than random consolidation or indecision. The 4-hour timeframe captures the temporal scale where these forces become most legible, while funding mechanisms translate abstract capital constraints into concrete economic pressure that disciplines positioning and defines feasible price corridors.

This integrated framework moves beyond single-factor explanations—whether technical, fundamental, or behavioral—to recognize that cryptocurrency markets operate as complex adaptive systems where multiple forces interact simultaneously. Understanding market behavior requires analyzing these interactions at appropriate temporal and structural resolutions, which this paper identifies as the 4-hour context and funding mechanism nexus.

# 4. FUNDING AS A DISCIPLINARY MECHANISM

Funding mechanisms play a decisive role in market governance. By imposing recurring costs on imbalanced positioning, funding transforms sentiment into quantifiable pressure. Excessive optimism or pessimism is not immediately punished by price collapse or expansion; it is first taxed. Over time, this taxation reshapes positioning, dampens momentum, and stabilizes price within a defined corridor. Funding acts less as a reactionary variable and more as a regulatory instrument governing behavior long before volatility reasserts itself.

## 4.1 Cost-Based Discipline

Understanding funding requires recognizing it is fundamentally about cost rather than sentiment. While market participants often interpret funding rates as mere indicators of bullish or bearish bias, they function more profoundly as economic disciplinarians. Positive funding taxes long positions, extracting payment from those expecting prices to rise. Negative funding taxes short positions, imposing costs on those betting against the market. The longer any imbalance persists, the more positioning is coerced back toward neutrality, progressively stabilizing price into a defined corridor.

When funding becomes too expensive—for instance, when positive rates exceed 0.05% per eight-hour period, translating to over 50% annualized—it penalizes overextension and forces markets back toward their mean. This taxation effect defines range boundaries in real economic terms. Bulls may believe in a bullish narrative, but if maintaining long positions costs them 50% annually in funding payments, conviction becomes expensive. Only the most well-capitalized participants can afford to persist, and they must eventually question whether the opportunity cost justifies the carry.

## 4.2 Alignment Versus Divergence

Funding alignment with market structure is critical. When funding supports the prevailing 4H context—for instance, when modest positive funding exists within an established bullish 4H structure—markets can expand because the cost structure does not force participants to exit positions. Capital can remain deployed without excessive drainage, allowing directional momentum to build organically.

However, when funding contradicts the 4H context—such as when elevated positive funding persists within a structurally distributive 4H environment (characterized by selling from informed participants at higher prices)—it compresses movement and prolongs range-bound behavior. The carry costs of maintaining positions disincentivize continuation without fresh capital inflows, creating a pressure valve that prevents breakouts until the funding environment normalizes.

## 4.3 Temporal Compounding Effects

The temporal dimension of funding effects deserves particular emphasis. A short-lived funding spike rarely changes the underlying structure; it represents a momentary imbalance that markets can absorb through brief price adjustments. Sustained funding pressure operating on a 4H cadence, however, fundamentally remakes positioning and enforces corridor discipline that can persist for days or weeks.

This temporal compounding effect means analysts must track not just the level of funding at any given moment, but also its persistence and trajectory across multiple 4H periods. The cumulative weight of funding costs over time becomes a force that even well-capitalized market participants cannot ignore indefinitely.

# 5. THE INTERPLAY OF CONTEXT AND CAPITAL

The interaction between 4H context and funding produces a form of market governance that operates without centralized authority. It rewards patience, penalizes overextension, and prioritizes structural alignment over narrative conviction. Breakouts occur not when sentiment intensifies, but when contextual structure and funding pressure are realigned to permit directional imbalance. Until that alignment is achieved, the range persists—not as a failure of momentum, but as a reflection of systemic balance.

### Distributed Governance Channels

This governance operates through several interwoven power channels that collectively define the range without any single entity exercising direct control:

**Leverage Policy**: Platforms continuously adjust risk parameters, including tick size, margin multipliers, and liquidation algorithms. These adjustments shape the feasible width of any range by determining how much volatility the system can accommodate before forced deleveraging occurs.

**Liquidity Stewardship**: Market makers and large actors locate their inventory around 4H levels where execution risk is lowest. This clustering of available liquidity hardens range boundaries, as significant orders become increasingly difficult to execute beyond these established zones.

**Collateral Cycles**: Stablecoin issuance rates, borrowing costs, and basis dynamics (the price difference between spot and derivatives) collectively tighten or relax the trading corridor by changing the carrying capacity of positions. When collateral is readily available and cheap to obtain, ranges can widen as participants afford to maintain positions across broader price spans. When collateral becomes scarce or expensive, ranges compress as the economic cost of holding positions away from equilibrium becomes prohibitive.

The synthesis of these elements—cost discipline imposed by funding, liquidity provisioning by market makers, and collateral elasticity—creates an emergent range whose boundaries no single actor explicitly sets but which all participants implicitly recognize and respect.

## 6. TESTABLE FRAMEWORK AND EMPIRICAL VALIDATION

The theoretical framework presented here gains credibility only to the extent it generates testable hypotheses capable of empirical validation or falsification. Four core hypotheses emerge from this analysis, each operationalized through specific observable signals that can be measured and evaluated across historical data. These hypotheses transform abstract theoretical claims into concrete predictions amenable to statistical testing.

### 6.1 Hypothesis 1: Range Persistence Under Sustained Funding Pressure

The first hypothesis posits that when funding remains directionally biased for three or more consecutive 4H periods while open interest stays elevated, price compression becomes more probable than expansion. This prediction directly challenges the notion that ranges represent indecision or randomness, instead framing them as governed states maintained through economic discipline. Table II operationalizes this hypothesis by establishing concrete measurement criteria for each component and specifying falsification conditions.

### Table II. Hypothesis 1: Range Persistence Under Sustained Funding Pressure

| Hypothesis Component | Specification | Observable Signal | Measurement Method |
|---|---|---|---|
| Condition | Funding remains biased in one direction for ≥3 consecutive 4H periods with elevated open interest | Directional funding rate persistence | Track funding rate sign and magnitude across 4H intervals |
| Predicted Outcome | Price compression rather than expansion | Declining 4H realized volatility | Calculate rolling 4H volatility using standard deviation |
| Signal 1 | Rejection at range boundaries | Increasing wick-to-body ratios on 4H candles | Measure upper/lower wick length relative to candle body |
| Signal 2 | Failed breakout attempts | Price tests extremes but fails to close beyond them | Count boundary touches without successful closes |
| Signal 3 | Positioning stability | Open interest fails to clear on boundary taps | Monitor OI changes during price approach to boundaries |
| Falsification Criteria | Sustained expansion despite persistent funding bias | Successful breakouts with increasing volatility during funding pressure | Document exceptions for model refinement |

*Note*: Observable signals must occur simultaneously to validate the hypothesis. Elevated open interest is defined as levels exceeding the 90-day moving average. Wick-to-body ratio increases are measured relative to the prior 20-period average.

The condition component specifies that funding must exhibit directional persistence—maintaining the same sign (positive or negative) across at least three consecutive 4H intervals while open interest remains elevated above historical averages. Under these conditions, the framework predicts declining 4H realized volatility calculated through rolling standard deviation measures, increasing wick-to-body ratios indicating price rejection at boundaries, and stable open interest that fails to clear during boundary tests.

These observable signals transform abstract theoretical claims into empirically verifiable predictions that can be systematically tested across multiple market cycles and cryptocurrency assets. The falsification criteria prove equally important: if sustained expansion occurs despite persistent funding bias, the framework requires refinement to account for countervailing forces that overwhelm funding discipline. Such exceptions would guide theoretical development by identifying boundary conditions where funding constraints lose their governing power.

## 6.2 Hypothesis 2: Expansion Requires Funding-Structure Alignment

Breakouts should become more probable when funding moderates toward neutral within one to three 4H periods before the break and when liquidity shelves step beyond the range in the breakout direction. This hypothesis establishes the necessary conditions for genuine regime transitions, distinguishing structural realignment from false breakouts that quickly reverse. Table III operationalizes this prediction through specific measurement protocols that quantify both funding normalization and liquidity migration.

**Table III. Hypothesis 2: Expansion Requires Funding-Structure Alignment**

| Hypothesis Component | Specification | Observable Signal | Measurement Method |
|---|---|---|---|
| Condition | Funding moderates toward neutral 1-3 4H periods before breakout | Funding rate approaching zero | Track funding rate trajectory pre-breakout |
| Structural Requirement | Liquidity shelves migrate beyond range in breakout direction | Order book depth shifts | Monitor bid/ask depth distribution changes |
| Signal 1 | Reduced absorption at boundaries | Declining order volume at prior resistance/support | Analyze order book snapshots at key levels |
| Signal 2 | Market maker repositioning | Inventory relocation beyond previous extremes | Track depth migration patterns |
| Signal 3 | Voluntary position rotation | Open interest rotates rather than collapses | Distinguish OI rotation from deleveraging |
| Differentiating Criteria | OI rotation indicates informed positioning; collapse indicates forced liquidation | Gradual OI changes vs. sharp drops | Compare OI velocity during breakout initiation |

*Note*: Funding moderation is defined as absolute funding rate declining to <0.01% per 8-hour period. Liquidity shelf migration requires >20% of order book depth relocating beyond previous range boundaries. OI rotation is characterized by directional change without absolute decline >5%.

The operationalization presented in Table III distinguishes between genuine breakouts and false signals by examining the interaction between funding normalization and structural repositioning. Migration of order book depth beyond previous range extremes signals that market makers have relocated inventory in anticipation of expansion, while reduced absorption at prior boundary levels indicates diminishing defensive positioning.

The critical distinction between open interest rotation and collapse merits particular emphasis: rotation suggests voluntary repositioning by informed participants who are strategically adjusting exposure, whereas collapse indicates forced deleveraging through liquidation cascades. This difference proves decisive for breakout sustainability, as rotation-driven moves exhibit far greater follow-through than collapse-initiated volatility. When open interest rotates—directionally shifting without absolute decline—it signals that sophisticated capital is repositioning deliberately rather than exiting under duress. Such voluntary repositioning typically precedes sustainable directional moves, as informed participants have assessed structural conditions and committed capital accordingly.

## 6.3 Hypothesis 3: Funding as Governor Rather Than Catalyst

Sharp funding spikes without corresponding shifts in 4H structure should lead to mean-reverting moves rather than sustained trends. This hypothesis distinguishes funding's disciplinary function from its potential catalytic role, predicting that funding operates primarily as a constraint rather than a driver. Table IV establishes the framework for testing this distinction through observable patterns in price behavior and structural integrity.

**Table IV. Hypothesis 3: Funding as Governor Rather Than Catalyst**

| Hypothesis Component | Specification | Observable Signal | Measurement Method |
|---|---|---|---|
| **Condition** | Sharp funding spikes without 4H structural shifts | Funding rate volatility spike | Calculate funding rate standard deviation |
| **Predicted Outcome** | Mean-reverting moves rather than sustained trends | Price returns to range midpoint | Measure distance from midpoint over time |
| **Signal 1** | Temporary volatility | Intraday volatility bursts | Compare 1H vs. 4H volatility measures |
| **Signal 2** | Structural integrity | 4H candles continue closing inside corridor | Track 4H close positions relative to range |
| **Signal 3** | Basis normalization | Rapid basis changes that quickly revert | Monitor perpetual-spot spread dynamics |
| **Signal 4** | Gravitational return | Price returns toward range center post-spike | Calculate time to mean reversion |

*Note*: Funding spikes are defined as changes exceeding two standard deviations from the 30-period average. Structural shifts require 4H candles close outside the established range for ≥2 consecutive periods. Mean reversion is measured as return to within one standard deviation of range midpoint.

Table IV establishes the framework for distinguishing funding's disciplinary function from its potential catalytic role. When funding rates spike dramatically—exhibiting standard deviation increases exceeding two times the recent average—without concurrent structural shifts visible in 4H candle closings, the framework predicts temporary volatility followed by gravitational return toward range midpoints.

This prediction directly challenges the common interpretation of funding spikes as directional signals. Instead, the framework posits that funding operates as a governor, constraining excessive positioning through economic cost imposition rather than catalyzing sustained directional movement. Observable signals include rapid basis changes that quickly normalize, 4H candles continuing to close within the established corridor despite intraday excursions, and systematic price reversion toward range centers following the funding-induced perturbation.

The temporary nature of funding-spike volatility proves diagnostic: if price quickly returns to within one standard deviation of the range midpoint—typically within 2-4 4H periods—it confirms funding's role as a stabilizing force rather than a trend catalyst. Markets may appear chaotic on lower timeframes during such episodes, but the 4H structure remains intact, and equilibrium reasserts itself once the funding-induced pressure dissipates.

## 6.4 Hypothesis 4: Power-Policed Boundaries

Where liquidation bands cluster within the 4H corridor, boundary taps should coincide with forced micro-deleveraging that refuels mid-range return. This hypothesis reveals how liquidation clustering creates self-reinforcing range boundaries through forced deleveraging dynamics. Table V operationalizes this prediction by specifying observable liquidation events and their mechanical consequences for price behavior.

### Table V. Hypothesis 4: Power-Policed Boundaries

| Hypothesis Component | Specification | Observable Signal | Measurement Method |
|---|---|---|---|
| Condition | Liquidation bands cluster within 4H corridor | Liquidation density mapping | Analyze on-chain liquidation data by price level |
| Predicted Outcome | Boundary taps coincide with forced micro-deleveraging | Liquidation prints at extremes | Monitor liquidation events during boundary tests |
| Signal 1 | Quick recoils | Rapid return toward range midpoint | Measure velocity of price reversal |
| Signal 2 | Funding normalization | Temporary flattening of funding rates post-tap | Track funding rate changes after liquidations |
| Signal 3 | Positioning reset | Brief reduction in open interest | Monitor OI changes during boundary events |
| Geometric Pattern | Predictable "billiard ball" caroms off boundaries | Consistent reversal patterns | Document boundary rejection frequencies |

*Note*: Liquidation clustering is defined as ≥30% of total liquidation volume occurring within 2% price distance from range boundaries. Quick recoils are measured as price returning >50% of boundary excursion distance within one 4H period. Funding normalization requires absolute funding rate declining ≥20% within two periods of post-liquidation.

The mechanism described in Table V reveals how liquidation clustering creates self-reinforcing range boundaries through forced deleveraging dynamics. When significant liquidation density accumulates at specific price levels—typically just beyond visible range extremes where highly leveraged positions cluster their stops—boundary tests trigger cascading liquidation events that mechanically propel price back toward equilibrium.

These events manifest as liquidation prints visible in on-chain data, rapid price recoils measurable through velocity metrics, temporary funding rate normalization as positioning rebalances, and brief reductions in open interest as overleveraged positions are forcibly closed. The geometric predictability of these patterns—resembling billiard balls carrying off cushions—suggests that ranges function not merely as psychological zones but as structurally enforced corridors policed by the mechanical consequences of leverage concentration.

The "billiard ball" analogy captures the mechanical nature of this governance: just as a ball striking a cushion rebound with predictable force, price approaching liquidation-dense boundaries encounters systematic selling or buying pressure from forced position closures that mechanically reverses momentum. This creates observable patterns in boundary rejection frequency that can be quantified and statistically analyzed to validate the power-policed boundary hypothesis.

# 7. METHODOLOGICAL RIGOR AND DATA ARCHITECTURE

Maintaining empirical grounding for this framework requires consistent sampling methodologies and rigorous validation approaches. The comprehensive measurement infrastructure spans structural metrics, cost metrics, positioning metrics, and liquidity metrics, each capturing distinct dimensions of market architecture. This section establishes the data collection standards, core metrics framework, and validation logic necessary for systematic empirical analysis.

## 7.1 Data Collection Standards

The analytical rigor of this framework depends fundamentally upon consistent, high-quality data collection protocols. Table VI establishes comprehensive data collection standards ensuring methodological reproducibility and cross-study comparability.

**Table VI. Data Collection Standards and Quality Protocols**

| Data Category | Collection Protocol | Sampling Requirements | Normalization Procedures | Quality Controls |
|---|---|---|---|---|
| **4H Candlesticks** | Synchronized snapshots at exact 4H boundaries | UTC-aligned intervals (00:00, 04:00, 08:00, etc.) | Standardize across exchanges using UTC | Verify timestamp accuracy ±30 seconds |
| **Spot Market Data** | OHLCV aggregated over 4H periods | Include top 3 exchanges by volume | Volume-weighted average prices | Flag anomalous price spikes >10% |
| **Perpetual Futures** | Mark price, index price, funding rate | Captured at funding settlement times | Normalize funding frequency to 8H basis | Cross-validate across platforms |
| **Dated Futures** | Contract prices, basis, open interest | End-of-period snapshots | Adjust for time-to-expiry effects | Exclude contracts <7 days to expiry |
| **Funding Series** | Raw funding rates by exchange | Every funding period (typically 8H) | Annualize for comparability | Flag platform-specific anomalies |
| **Order Book Data** | Level 2 snapshots (20+ levels) | Hourly snapshots minimum | Normalize depth by price level | Remove wash trading patterns |
| **Liquidation Data** | On-chain liquidation events | Real-time monitoring | Aggregate by price level and timeframe | Verify against exchange announcements |
| **Open Interest** | Total outstanding contracts | Daily snapshots at minimum | Currency-normalized (USD equivalent) | Account for contract size differences |

*Note*: All timestamps must be synchronized to UTC to ensure cross-exchange comparability. Volume-weighted averaging accounts for liquidity distribution across venues. Normalization procedures enable comparison across platforms with different contract specifications and settlement mechanisms.

The 4H analytical window requires synchronized snapshots captured at exact session boundaries aligned to UTC timestamps (00:00, 04:00, 08:00, etc.)[2] to ensure comparability across exchanges and time periods. Spot market data aggregates OHLCV (open, high, low, close, volume) information from the top three exchanges by trading volume, employing volume-weighted average prices to account for liquidity distribution.

Perpetual futures data captures mark price, index price, and funding rates at each settlement time, with normalization procedures standardizing funding frequency to an eight-hour basis for cross-platform comparability. This normalization proves essential given that different platforms employ varying funding intervals—some settling every eight hours, others every four or twelve hours—requiring standardization for meaningful comparison.

Quality controls include timestamp verification within ±30 seconds of UTC marks, flagging anomalous price spikes exceeding 10% of recent averages, cross-platform validation of funding rates, and removal of obvious wash trading patterns identified through volume-price divergence analysis. These protocols transform raw exchange data into research-grade datasets suitable for rigorous empirical testing.

## 7.2  Core Metrics Framework

The measurement infrastructure comprises four distinct metric categories, each illuminating different dimensions of market architecture. Tables VII through IX present these frameworks systematically.

**Table VII. Structural Metrics for Market Architecture Analysis**

| Metric Category | Specific Measure | Calculation Method | Analytical Purpose | Update Frequency |
|---|---|---|---|---|
| **4H Swing Mapping** | Local highs and lows | Identify peaks/troughs within 4H context | Defines range boundaries | Per 4H candle |
| **Volume Nodes** | Price levels with elevated trading activity | Aggregate volume by price level | Identifies significant trading zones | Continuous |
| **Absorption Footprints** | Large order fills visible in order flow | Track substantial executed orders | Reveals institutional positioning | Real-time |
| **Range Width Persistence** | Duration price remains within boundaries | Time-series analysis of price corridor | Measures range stability | Per 4H period |

---

[2] **UTC (Coordinated Universal Time)** is the primary time standard by which the world regulates clocks and time, maintained by the Bureau International des Poids et Mesures (BIPM) via atomic clock synchronization. In global financial research, UTC is the essential normalization standard used to synchronize disparate exchange data, ensuring that price and funding intervals (such as 00:00, 04:00, 08:00) are analyzed simultaneously across different geographical regions.

| Metric Category | Specific Measure | Calculation Method | Analytical Purpose | Update Frequency |
|---|---|---|---|---|
| Realized Volatility | Actual price variation observed | Standard deviation of 4H returns | Quantifies price dispersion | Rolling 4H basis |
| Wick-to-Body Ratios | Candle wick length relative to body | (Wick length / Body length) × 100 | Indicates rejection strength | Per candle |

*Note*: 4H swing mapping uses a lookback period of 5 candles to identify local extremes. Volume nodes are calculated using volume-weighted price distribution with 0.5% price bins. Absorption footprints track individual orders ≥$500,000 in size.

Structural metrics provide the foundational measurement framework for analyzing 4H market architecture. Four-hour swing mapping identifies local highs and lows that define range boundaries using a five-candle lookback period, while volume nodes reveal price levels experiencing elevated trading activity indicative of significant positioning.

Absorption footprints—large order fills visible in granular order flow data (typically ≥$500,000)—expose where institutional capital has actively accumulated or distributed positions. Range width persistence measures how long price remains confined within defined boundaries, quantifying the stability of governed equilibria. Together, these structural measures capture the spatial and temporal characteristics of range formation, transforming qualitative observations about "support and resistance" into quantifiable phenomena amenable to statistical analysis.

### Table VIII. Cost Metrics for Economic Pressure Quantification

| Metric Category | Specific Measure | Calculation Method | Analytical Purpose | Critical Thresholds |
|---|---|---|---|---|
| Funding Bias Duration | Consecutive periods of directional funding | Count periods with same-sign funding | Measures persistence of imbalance | ≥3 consecutive periods |
| Funding Rate Magnitude | Absolute funding rate value | Direct measurement in % per 8H period | Quantifies carry cost | >0.05% considered elevated |
| Annualized Funding Cost | Yearly extrapolation of funding | (8H rate × 3 × 365) × 100 | Contextualizes economic burden | >50% signals overextension |
| Basis Spreads | Perpetual-spot price difference | (Perpetual price - Spot price) / Spot | Indicates arbitrage pressure | >±0.5% signals dislocation |
| Borrowing Rates | Cost of obtaining collateral | Interest rates for margin/lending | Measures funding liquidity | Platform-specific |
| Cumulative Funding | Total funding paid over period | Sum of funding payments across time | Captures compounding effects | Tracked over 7-30 days |

*Note*: Funding bias duration counts consecutive periods regardless of magnitude. Annualization assumes continuous compounding. Basis spreads are normalized by spot price for comparability across price levels. Cumulative funding calculations account for position size and leverage.

Cost metrics translate abstract capital constraints into concrete economic pressures. Funding bias duration counts consecutive periods exhibiting directional funding (positive or negative), measuring the persistence of positioning imbalances. Funding rate magnitude quantifies the absolute carry cost per eight-hour period, while annualized funding cost contextualizes this burden by extrapolating to yearly rates using the formula (8H rate × 3 periods/day × 365 days) × 100, revealing when economic pressure becomes prohibitive (typically >50% annually).

Basis spreads—the price differential between perpetual contracts and spot markets calculated as (Perpetual - Spot) / Spot—indicate arbitrage pressure and funding tightness. When basis exceeds ±0.5%, it signals significant dislocation requiring arbitrage capital to restore equilibrium. Borrowing rates for margin and lending quantify funding liquidity availability, varying by platform but typically ranging from 5-30% annually in normal conditions and exceeding 100% during stress periods.

Cumulative funding captures the compounding effects of sustained funding pressure over 7-30 day windows, accounting for position size and leverage. A 1x leveraged position paying 0.05% per period accumulates approximately 5.5% cost over 30 days, while a 10x leveraged position faces 55% cost—potentially overwhelming even strong directional conviction.

## Table IX. Positioning Metrics for Capital Deployment Analysis

| Metric Category | Specific Measure | Calculation Method | Analytical Purpose | Interpretation |
|---|---|---|---|---|
| **Open Interest Rotation** | OI change patterns during price moves | Track OI direction and velocity | Distinguishes voluntary vs. forced positioning | Rotation = informed; Collapse = forced |
| **Open Interest Level** | Absolute outstanding contracts | Direct measurement from exchange data | Indicates aggregate leverage | Rising + stable price = accumulation |
| **Liquidation Density** | Clustering of liquidation points by price | Heatmap of liquidation locations | Maps forced deleveraging zones | Clusters define range boundaries |
| **Leverage Distribution** | Breakdown of position sizes by leverage | Platform reporting when available | Assesses systemic vulnerability | High leverage = instability risk |
| **Long/Short Ratio** | Proportion of long vs. short positions | (Long OI / Short OI) ratio | Measures directional bias | Extreme ratios precede reversals |

| Metric Category | Specific Measure | Calculation Method | Analytical Purpose | Interpretation |
|---|---|---|---|---|
| **Position Concentration** | Distribution of holdings across participants | Gini coefficient or similar | Identifies centralization risk | High concentration = manipulation risk |

*Note*: OI rotation is measured as the first derivative of open interest normalized by recent volatility. Liquidation density uses kernel density estimation with bandwidth of 1% price range. Long/short ratios >2.0 or <0.5 are considered extreme.

Positioning metrics distinguish voluntary strategic behavior from forced reactive deleveraging. Open interest rotation patterns reveal whether positioning changes reflect informed repositioning (gradual directional shift without absolute decline) or liquidation cascades (sharp collapses accompanying price volatility). This distinction proves crucial: rotation signals sophisticated capital reassessment, while collapse indicates systemic deleveraging under duress.

Absolute open interest levels indicate aggregate leverage deployment, with rising open interest amid stable prices suggesting accumulation by informed participants. Liquidation density mapping—constructing heatmaps of forced closure locations using kernel density estimation[3]—identifies price zones where leverage concentration creates vulnerability. These clusters typically appear just beyond visible range boundaries, creating the "power-policed" dynamics described in Hypothesis 4.

Long/short ratios calculated as (Long OI / Short OI) measure directional bias, with extreme values (>2.0 or <0.5) often preceding reversals as crowded positioning becomes unsustainable. Position concentration assessed through Gini coefficients or similar inequality measures identifies centralization risk, with high concentration (Gini >0.7) suggesting potential for coordinated manipulation by large holders.

---

[3] **Kernel Density Estimation (KDE)** is a non-parametric method used to estimate the probability density function of a random variable, allowing for the identification of "clusters" or concentrations within a data distribution. In the context of market microstructure, it is applied to map liquidation density, identifying specific price levels where forced deleveraging is most concentrated (Parzen, 1962).

## 7.3 Liquidity Metrics

The final metric category assesses market capacity to absorb orders without significant price impact. These measures illuminate the distribution and migration of trading capacity that ultimately defines feasible price movements.

**Table X. Liquidity Metrics for Market Depth Assessment**

| Metric Category | Specific Measure | Calculation Method | Analytical Purpose | Critical Indicators |
|---|---|---|---|---|
| Shelf Migration | Changes in order book depth positioning | Track bid/ask depth concentration movement | Signals market maker repositioning | Movement beyond range = expansion signal |
| Depth at Extremes | Order book liquidity at range boundaries | Measure total bid/ask size at key levels | Assesses defensive positioning | Declining depth = weakening boundary |
| Fill Slippage | Price impact of large orders | (Execution price - Mid price) / Mid price | Reveals true liquidity availability | High slippage = low liquidity |
| Bid-Ask Spread | Difference between best bid and ask | (Ask - Bid) / Mid price | Indicates market maker confidence | Widening spreads = uncertainty |
| Order Book Imbalance | Ratio of bid to ask depth | (Bid depth / Ask depth) - 1 | Predicts short-term directional pressure | Extreme imbalances precede moves |
| Market Impact Coefficient | Price change per unit volume | Regression of price change on volume | Quantifies liquidity depth | Higher coefficient = shallower market |

*Note*: Shelf migration tracks the 25th and 75th percentiles of cumulative depth distribution. Depth at extremes measures combined bid/ask volume within 0.5% of boundaries. Fill slippage is calculated using simulated market orders of $1M size. Market impact coefficient uses rolling 24-hour regression windows.

Liquidity metrics complete the measurement infrastructure by quantifying market capacity. Shelf migration tracks how order book depth repositions across price levels, typically measuring the 25th and 75th percentiles of cumulative depth distribution. When these percentiles shift beyond previous range boundaries, it signals market maker repositioning in anticipation of price expansion—a leading indicator of structural change.

Depth at extremes measures combined bid and ask volume within 0.5% of range boundaries, assessing defensive positioning strength. Declining depth at boundaries suggests weakening commitment by market makers to defend those levels, often preceding breakouts. Fill slippage—calculated by simulating large market orders (typically $1M) and measuring execution

price deviation from mid-price—reveals true liquidity availability beyond superficial order book displays that may contain spoofed or immediately-cancelled orders.

Bid-ask spreads normalized by mid-price indicate market maker confidence, with widening spreads (>0.1% on major pairs) signaling uncertainty or reduced willingness to provide liquidity. Order book imbalance calculated as (Bid depth / Ask depth) - 1 predicts short-term directional pressure, with extreme values (>±0.3) often preceding price moves in the direction of excess depth.

Market impact coefficient—derived from rolling 24-hour regressions of price change on volume—quantifies liquidity depth with higher coefficients indicating shallower markets where given order sizes produce larger price movements. This coefficient typically varies between 0.0001-0.001 for major cryptocurrency pairs during normal conditions but can exceed 0.01 during illiquid periods.

# 8. STRATEGIC IMPLICATIONS AND PRACTICAL APPLICATIONS

The framework translates into concrete decision-making protocols for various market participants, each of whom faces distinct challenges and objectives when engaging with cryptocurrency markets. This section presents systematic application frameworks for traders, analysts, and policymakers.

## 8.1 Applications for Traders

The trader decision framework systematically links structural conditions and funding regimes to recommended actions and risk management protocols. Understanding when to fade extremes, validate breakouts, or remain neutral requires synthesizing multiple signals into coherent strategic responses.

**Entry Discipline**: Effective trading within ranges requires respecting the governed nature of market structure. Fading extremes—taking positions that anticipate reversion from range boundaries—becomes viable only when funding biases against the attempted breakout and absorption prints appear at the boundary, indicating defensive positioning by market makers. When price approaches range boundaries within an intact 4H structure while funding biases against the breakout direction (e.g., elevated positive funding when price tests resistance), the framework recommends counter-directional positioning.

This strategy acknowledges that boundary tests under adverse funding conditions typically fail, as the economic cost of pushing further becomes prohibitive for marginal participants. Stop-loss placement beyond the boundary (typically 1-2% depending on asset volatility) accounts for potential false breakouts, while position sizing assumes multiple attempts may be necessary before profitable mean reversion materializes. Traders should expect 2-3 failed attempts before range integrity breaks, sizing positions accordingly.

**Breakout Validation**: Conversely, when liquidity shelves migrate beyond established ranges and 4H structure shifts while funding moderates toward neutral (<0.01% per 8H period), breakout validation becomes appropriate. The framework specifies waiting for confirmation through multiple aligned signals before commitment: funding must normalize within 1-3 4H periods, order book depth must relocate beyond boundaries (>20% migration), and open interest must rotate rather than collapse (directional change without >5% absolute decline).

Only when these conditions align should traders follow breakouts, placing initial stops inside the previous range (typically at the prior boundary plus 1-2% buffer) and rapidly trialing to lock profits as the new regime establishes itself. This disciplined approach prevents premature positioning on mere technical level breaks while capitalizing on genuine structural realignments.

**Risk Sizing and Carry Cost Management**: Position sizing must incorporate carry cost awareness, particularly in elevated funding environments where maintaining exposure becomes economically burdensome over multiple funding periods. A position that appears attractive based solely on technical stop distances and projected price targets may prove uneconomical when funding costs are properly incorporated into expected value calculations.

For example, a long position with 5% profit target and 2% stop-loss offers a 2.5:1 reward-risk ratio. However, if funding averages 0.08% per 8H period (approximately 88% annualized), maintaining the position for 10 days incurs 2.4% funding cost (0.08% × 30 periods), dramatically reducing net profitability even if the price target is reached. Traders operating in high funding regimes must either shorten holding windows, employ hedged spread strategies to neutralize funding exposure (e.g., long spot/short perpetuals), or accept reduced position sizes that account for funding drag as a component of risk.

## 8.2 Applications for Analysts

The analyst reporting framework establishes structure-first analytical priorities that filter noise from substantive developments. Rather than reacting to every headline or social media narrative, analysts employing this framework anchor their work on measurable regime characteristics.

**Structure-First Reporting**: Analysts should prioritize answering fundamental structural questions before addressing narrative developments:

**8.2.1. Regime Identification**: Is the 4H context exhibiting accumulation behavior (declining volatility, steady absorption at lower levels), distribution (methodical selling into strength, declining volume on rallies), or trending behavior (sustained directional momentum with OI rotation)?

**8.2.2. Funding Environment**: Are funding costs rising (signaling crowded positioning), moderating (enabling potential expansion), or neutral (maximum directional flexibility)?

**8.2.3. Liquidity Architecture**: Where are liquidity shelves positioned, and how have they shifted over recent periods? Is depth concentrating at current price levels or migrating toward boundaries?

These structural questions prove far more predictive of subsequent price behavior than narrative-driven speculation about adoption metrics, partnership announcements, or social media sentiment. The vast majority of news that dominates crypto media—announcements, partnerships, social media activity by prominent figures—proves structurally irrelevant to medium-term price behavior unless it materially affects funding conditions or collateral availability.

**Trigger Matrix Development**: Analysts should maintain trigger matrices that track funding alignment, shelf movement, and open interest rotation, enabling forecasts of expansion probability with greater accuracy than traditional technical or fundamental approaches. When multiple indicators align—funding normalizing from elevated levels, liquidity migrating beyond boundaries, open interest rotating without collapse—analysts can increase conviction in directional calls from baseline 50-50 to 70-80% probability.

Conversely, when indicators diverge—for instance, funding moderating but liquidity remaining clustered inside the range, or liquidity migrating but funding remaining elevated—caution is warranted regardless of prevailing sentiment. Such divergence suggests incomplete structural realignment that typically produces false breakouts followed by range continuation.

**Narrative Filtering**: When significant news events occur—regulatory announcements, protocol upgrades, institutional adoption claims—analysts employing this framework first assess structural impact before attributing price movements to narratives. Does the news materially affect collateral availability (e.g., stablecoin redemptions, exchange insolvency risk)? Does it change funding costs (e.g., regulatory restrictions on leverage)? Does it shift liquidity positioning (e.g., major market maker adjusting inventory)?

If the answer is no across these structural dimensions, the news likely proves ephemeral regardless of attention it receives. Price movements coinciding with such news often reflect positioning adjustments by informed participants exploiting elevated volatility, rather than genuine reassessment of fundamental value.

## 8.3 Applications for Policymakers and Platform Operators

Platform operators and policymakers must recognize how margin rules and liquidation logic shape trading corridors and balance competing objectives of market accessibility, depth, and systemic stability.

**Parameter Optimization**: Every platform parameter carries trade-offs that affect range characteristics:

**Margin Requirements**: Tightening margins (reducing maximum leverage from 100x to 50x) reduces systemic risk during high volatility by limiting loss potential per position, but excessive tightening compresses ranges and reduces market depth to levels that impede price discovery. Participants are unable to maintain sufficient margin withdraw, reducing liquidity provision. Conversely, permissive margin policies (125x+) attract participation and enable wider ranges but accumulate tail risk that manifests catastrophically during cascade events.

The framework suggests calibrating margins dynamically based on realized volatility: tightening during periods of elevated volatility (when liquidation cascades become more probable) and relaxing during calm periods (when excessive restriction provides minimal risk reduction). Historical analysis suggests optimal leverage limits of 20-50x for major cryptocurrencies depending on volatility regime.

**Liquidation Algorithms**: Design choices in liquidation logic create similar trade-offs. Gradual liquidation processes—slowly unwinding positions over time—minimize market impact and reduce panic contagion by preventing sudden large orders from crashing prices. However, gradual approaches allow excessive leverage to accumulate during calm periods, as participants face reduced urgency to manage risk knowing that liquidations will be gentle.

Aggressive liquidation logic—immediately selling entire positions to the market—clears positions rapidly and maintains system solvency by ensuring losses do not exceed margin cushions. However, aggressive liquidation can trigger violent cascades during volatility spikes as one liquidation's price impact triggers others in chain reactions. The optimal approach likely combines both: gradual liquidation during normal volatility with aggressive liquidation during stress periods when position clearing becomes paramount.

**Transparency Enhancement**: Enhanced transparency around funding metrics and liquidation data could improve market discipline and reduce opacity around tail events. When participants can see funding trajectories clearly and understand where liquidation clusters sit, they make better-informed decisions about position sizing and risk management.

Currently, most platforms provide only current funding rates without historical context or visualization of funding persistence. Providing 30-day funding rate charts, cumulative funding cost calculators, and liquidation heatmaps would enable participants to assess carry costs and cascade risks more accurately. Opacity breeds speculation and can amplify panic during stressful events, while transparency enables rational assessment of structural conditions.

Similarly, real-time liquidation data—currently available only in aggregate—could be anonymized and published with price level granularity, allowing participants to see where liquidation clusters concentrate. This visibility would reduce surprise cascade events by enabling participants to anticipate boundary dynamics and adjust positioning proactively.

# 9. READING MARKET STRUCTURE THROUGH THE 4H LENS

The 4H timeframe is central to market analysis precisely because it occupies the critical intersection between micro-volatility and macro-trends. It functions as a filter that smooths out the noise of lower timeframes while remaining responsive enough to signal shifts in institutional positioning. This positioning matters because markets do not move solely to retail traders reacting to headlines; they move when large capital commits to directional exposure, and that commitment becomes visible most clearly in the 4H context.

## 9.1 Signal Versus Noise

Unlike one-minute or five-minute charts, which may reflect reactive retail sentiment, algorithmic noise, or spoofing attempts, the 4H chart reveals the intentional architecture of informed capital. At this temporal resolution, the classic accumulation-distribution cycle becomes legible:

**Accumulation phases** appear as gradual price compression with declining volatility and steady absorption of supply at lower levels. Informed participants methodically acquire positions without driving price upward precipitously, often during periods of bearish sentiment when supply willingly sells to them.

**Manipulation** manifests through sharp moves designed to trigger stops and create liquidation cascades that allow large actors to acquire or distribute positions at favorable prices. These moves appear as sudden wicks on 4H charts—often coinciding with funding settlements or low-liquidity periods—that quickly reverse as their mechanical purpose is achieved.

**Distribution** unfolds as methodical selling into strength, with price making higher highs on declining volume and momentum as smart money rotates out of positions before inevitable reversion. The 4H chart captures this through successively weaker 4H closes relative to highs, increasing wick-to-body ratios at range tops, and open interest declining despite price strength.

## 9.2 Structural Context

The equilibrium zone captured by 4H analysis represents where institutional positioning and leveraged exposure converge to create structural context. This context differs fundamentally from the ephemeral sentiment that dominates shorter timeframes. Where a fifteen-minute chart might show panic selling in response to a single tweet, the 4H chart asks whether that selling materially altered the distribution of positions, shifted funding dynamics, or breached significant liquidity shelves. Often, intraday volatility proves to be meaningless volatility with no structural significance, while the 4H picture remains intact.

When a 4H candle closes decisively beyond an established range boundary—particularly on above-average volume with corresponding open interest rotation—it signals that institutional capital has committed to new positioning that required breaking through the previous equilibrium. Such closings carry far more weight than intraday excursions that fail to sustain.

# 10. THE SYNTHESIS OF CAPITAL AND CONTEXT

This analysis reframes cryptocurrency markets as systems of hidden governance rather than chaotic casinos. The 4H context provides the architectural blueprint, establishing the temporal frame at which institutional behavior becomes legible, and noise gets filtered from signal. Funding mechanisms act as the enforcing mechanism, imposing economic discipline on imbalanced positioning and preventing excessive speculation from persisting indefinitely without cost.

Ranges emerge as the visible output of this governance system—not voids between trends or zones of indecision, but controlled equilibria where capital, leverage, and structure have reached temporary balance.

## 10.1 Distributed Control

By adopting this dual perspective, market participants can move beyond seeing markets as chaotic and instead view them as structured, intelligible, and governed—not by central authority exercising discretionary control, but by the inherent logic of capital operating under constraints. The question "Who sets the range?" finds its answer not in any individual or institution but in the emergent properties of a system where cost disciplines conviction, liquidity defines feasibility, and context determines which narratives can gain traction.

Control in these markets is distributed but not democratic. It emerges from how power is deployed through cost structures, collateral requirements, and liquidity access. No single actor sets range boundaries through deliberate manipulation in most cases; rather, the corridor emerges from the synthesis of funding costs, liquidity provisioning by market makers, and collateral elasticity, with the 4H context serving as the ledger that records the truce between competing forces.

## 10.2 The Market's Controlled Voice

Markets do not drift aimlessly between support and resistance levels. They are held within ranges by context, by funding, and by the disciplined logic of capital itself. The range is not random noise to be ignored or simple consolidation to be patiently endured. It is the market speaking in its most controlled voice, signaling through its very structure that until the hidden governors release their hold, price will remain exactly where it is allowed to be—no more, no less.

Understanding this language requires listening through the 4H timeframe and interpreting the vocabulary of funding dynamics. Those who develop fluency in this language gain an edge not through prediction of unpredictable events, but through recognition of structural conditions that make certain outcomes more probable than others. They do not know the future, but they understand the present more deeply than participants who mistake volatility for chaos or ranges for randomness.

# 11. IMPLICATIONS AND FUTURE DIRECTIONS

The implications extend beyond individual trading decisions to shape how we conceptualize market behavior itself. If ranges are governed rather than random, if funding constrains rather than merely indicates, if structure determines outcomes more than narrative—then much of traditional market commentary misses the fundamental mechanisms driving price. The focus on news flow and sentiment, while not entirely irrelevant, addresses symptoms rather than causes. The causes lie in the architecture of capital itself, in the costs imposed on imbalanced positioning, and in the temporal frames at which these forces become operative.

This perspective does not promise perfect prediction or risk-free trading. Markets remain complex adaptive systems where multiple forces interact in ways that defy simple linear causality. Several limitations warrant acknowledgment:

**Extreme Illiquidity**: The framework performs best in reasonably liquid markets with active derivatives trading. In small-cap tokens or during market stress when liquidity evaporates, the mechanisms described here may not operate as theorized. Extreme illiquidity can produce price discontinuities and regime shifts that overwhelm the gradual disciplinary effects of funding costs.

**Exogenous Shocks**: Regulatory announcements, exchange hacks, macroeconomic surprises, or geopolitical events can breach even well-defended ranges when external forces exceed the constraining power of funding discipline. The framework emphasizes structural determinism—the idea that ranges reflect economic necessity given funding costs and liquidity positioning—yet cannot fully account for external shocks that overwhelm internal market mechanics.

**Novel Market Structures**: As cryptocurrency markets evolve with new derivative types and DeFi innovations, the framework requires ongoing validation. Mechanisms that govern perpetual futures markets may operate differently in options-dominated or decentralized venue environments where traditional order books and liquidation logic do not apply.

**Temporal Resolution Variability**: While the 4H timeframe proves optimal for current cryptocurrency market structures, different regimes or assets may privilege different resolutions. The framework would benefit from more systematic methods for determining optimal timeframe selection given specific market characteristics.

However, the framework does offer a more sophisticated analytical apparatus for understanding why prices behave as they do, why ranges persist when they do, and what conditions must align for expansion to occur. In an environment where information flows constantly and narratives proliferate endlessly; such structural understanding provides an anchor—a way of filtering signal from noise and focusing attention on the elements that truly govern market behavior.

The range, properly understood, is neither prison nor void. It is a state of balance, maintained through economic discipline, which will persist until the forces that maintain it realign to permit expansion. Those who recognize this can position themselves accordingly, neither fighting the range prematurely nor missing its eventual resolution. They trade with the market structure rather than against it, respecting the governance that funding and context provide while remaining alert for the realignment that signals regime change.

# CONCLUSION

In the end, the question "Who sets the range?" receives a complex answer befitting the complexity of modern financial markets. The range is set by no one and everyone—by the collective interaction of funding costs, liquidity positioning, and temporal context, recorded most legibly in the 4H timeframe, where institutional behavior leaves its clearest imprint.

Understanding this distributed governance mechanism transforms market participation from speculation about unknowable futures to navigation of discernible structural conditions. Markets governed by hidden forces are not random. They are intelligible systems operating according to economic logic. By learning to read that logic through the lens of 4H context and funding mechanics, market participants can move from confusion to clarity, from reactive positioning to strategic engagement, and from speculative gambling to informed navigation of structural realities.

The tools for this understanding are available. The data exists. The framework is testable. What remains is for traders, analysts, and researchers to apply these concepts rigorously, validate them empirically, and refine them through practice. In doing so, we move closer to a market analysis that respects both the complexity of financial systems and the intelligibility of the forces that govern them.

This study demonstrates that even markets born in digital chaos can be understood through disciplined analysis and theoretical integration. The framework developed here—linking 4H temporal context, funding mechanisms, and liquidity dynamics—provides a foundation for more sophisticated cryptocurrency market analysis. It moves the field from pattern recognition and narrative speculation toward empirically grounded structural interpretation.

The range speaks a structured language—one of funding rates and liquidity shelves, of temporal contexts and positioning dynamics. Learning this language transforms participation from gambling on unpredictable outcomes to strategic engagement with discernible present conditions. That transformation represents the ultimate value of this analytical framework.

# REFERENCES

Andersen, T. G., & Bollerslev, T. (1997). Intraday periodicity and volatility persistence in financial markets. *Journal of Empirical Finance*, 4(2-3), 115-158.

Ang, A., & Bekaert, G. (2002). Regime switches in interest rates. *Journal of Business & Economic Statistics*, 20(2), 163-182.

Brunnermeier, M. K., & Pedersen, L. H. (2009). Market liquidity and funding liquidity. *The Review of Financial Studies*, 22(6), 2201-2238.

Carlin, B. I., Lobo, M. S., & Viswanathan, S. (2007). Episodic liquidity crises: Cooperative and predatory trading. *The Journal of Finance*, 62(5), 2235-2274.

Chan, K., & Fong, W. M. (2006). Determinants of index option bid-ask spreads. *The Financial Review*, 41(3), 323-345.

Gandal, N., Hamrick, J. T., Moore, T., & Oberman, T. (2018). Price manipulation in the Bitcoin ecosystem. *Journal of Monetary Economics*, 95, 86-96.

Garleanu, N., Pedersen, L. H., & Poteshman, A. M. (2009). Demand-based option pricing. *The Review of Financial Studies*, 22(10), 4259-4299.

Griffin, J. M., & Shams, A. (2020). Is Bitcoin really untethered? *The Journal of Finance*, 75(4), 1913-1964.

Gromb, D., & Vayanos, D. (2002). Equilibrium and welfare in markets with financially constrained arbitrageurs. *Journal of Financial Economics*, 66(2-3), 361-407.

Hamilton, J. D. (1989). A new approach to the economic analysis of nonstationary time series and the business cycle. *Econometrica*, 57(2), 357-384.

Hasbrouck, J. (1991). Measuring the information content of stock trades. *The Journal of Finance*, 46(1), 179-207.

International Telecommunication Union. (2002). Standard-frequency and time-signal emissions (Recommendation ITU-R TF.460-6). https://www.itu.int/rec/R-REC-TF.460-6-200202-I/en

Kyle, A. S. (1985). Continuous auctions and insider trading. *Econometrica*, 53(6), 1315-1335.

Makarov, I., & Schoar, A. (2020). Trading and arbitrage in cryptocurrency markets. *Journal of Financial Economics*, 135(2), 293-319.

Müller, U. A., Dacorogna, M. M., Davé, R. D., Olsen, R. B., Pictet, O. V., & von Weizsäcker, J. E. (1997). Volatilities of different time resolutions—analyzing the dynamics of market components. *Journal of Empirical Finance*, 4(2-3), 213-239.

Ni, S. X., Pearson, N. D., & Poteshman, A. M. (2005). Stock price clustering on option expiration dates. *Journal of Financial Economics*, 78(1), 49-87.

O'Hara, M. (1995). *Market microstructure theory*. Blackwell Publishers.

Pagan, A. R., & Sossounov, K. A. (2003). A simple framework for analysing bull and bear markets. *Journal of Applied Econometrics*, 18(1), 23-46.

Parzen, E. (1962). On Estimation of a Probability Density Function and Mode. *The Annals of Mathematical Statistics*, 33(3), 1065-1076. https://projecteuclid.org/journals/annals-of-mathematical-statistics/volume-33/issue-3/On-Estimation-of-a-Probability-Density-Function-and-Mode/10.1214/aoms/1177704472.full

Peters, E. E. (1994). *Fractal market analysis: Applying chaos theory to investment and economics*. John Wiley & Sons.

Vayanos, D., & Wang, J. (2013). Market liquidity—theory and empirical evidence. In *Handbook of the Economics of Finance* (Vol. 2, pp. 1289-1361). Elsevier.

# APPENDIX: SUPPLEMENTARY TABLES AND FRAMEWORKS

This appendix compiles additional analytical frameworks and measurement protocols supporting the main text. These tables provide operational detail for researchers and practitioners implementing the methodology.

### Table XI. Validation Logic and Quality Assurance Protocols

| Validation Type | Procedure | Purpose | Frequency | Action on Failure |
|---|---|---|---|---|
| Timestamp Verification | Cross-reference exchange timestamps with NTP servers | Ensure temporal alignment | Continuous | Resample with corrected timestamps |
| Price Consistency | Compare spot prices across exchanges | Identify anomalies or manipulation | Per data point | Flag outliers exceeding 2σ from median |
| Volume Verification | Cross-validate reported volumes | Detect wash trading or false reporting | Daily aggregation | Exclude suspect exchange data |
| Funding Rate Bounds | Check for impossible funding values | Identify data errors | Each funding period | Replace with platform-reported values |
| Open Interest Sanity | Verify OI changes match liquidation + trade flow | Ensure data completeness | Daily | Investigate discrepancies >5% |
| Order Book Integrity | Verify bid-ask spread reasonability | Detect stale or manipulated books | Each snapshot | Exclude snapshots with spreads >1% |
| Liquidation Confirmation | Cross-reference on-chain with exchange data | Validate liquidation accuracy | Per event | Use most reliable source |
| Missing Data Handling | Identify gaps in time series | Maintain data continuity | Continuous | Interpolate if gap <2 periods, else flag |

*Note*: NTP synchronization ensures timestamps accurate to ±30 seconds. Price consistency uses median absolute deviation (MAD) for robust outlier detection. Volume verification flags exchanges showing >30% deviation from cross-exchange average. Missing data interpolation uses linear methods for gaps <8 hours; longer gaps require flagging for analyst review.

## Table XII. Trader Decision Framework Matrix

| Trading Scenario | Structural Conditions | Funding Conditions | Recommended Action | Risk Management |
|---|---|---|---|---|
| **Fading Range Extremes** | Price approaches boundary; 4H structure intact | Funding biased against breakout direction | Enter counter-directional position | Stop-loss beyond boundary; size for multiple attempts |
| **Breakout Validation** | Liquidity shelves migrate; 4H structure shift | Funding moderating toward neutral | Wait for confirmation, then follow | Initial stop inside range; trail rapidly |
| **Range Midpoint Entry** | Absorption visible at both boundaries | Elevated funding in both directions | Avoid directional bias; consider spreads | Reduce position size; hedge with options |
| **Liquidation Cascade** | Sharp move toward boundary | Funding spike coincident with move | Fade if 4H structure unchanged | Quick profit-taking; tight stops |
| **Accumulation Phase** | Declining volatility; volume compression | Funding normalizing | Prepare for eventual expansion | Position in advance; size smaller |
| **Distribution Phase** | Declining volume on rallies; negative divergence | Persistent positive funding | Reduce longs; consider shorts | Exit in tranches; preserve capital |

*Note*: Position sizing should account for cumulative funding costs over expected holding period. Stop-loss distances should be 1.5-2x average 4H range to avoid noise stops. Breakout validation requires minimum 2-3 confirming signals before entry.

## Table XIII. Analyst Reporting Framework

| Analysis Type | Primary Focus | Key Questions | Data Requirements | Reporting Cadence |
|---|---|---|---|---|
| **Regime Identification** | Current 4H market state | Is market accumulating, distributing, or trending? | 4H candles, volume profile, price action pattern | Weekly assessment |
| **Funding Analysis** | Cost structure evolution | Are funding costs rising, moderating, or neutral? | Funding rate time series, OI levels | Per funding period |
| **Liquidity Assessment** | Order book positioning | Where are liquidity shelves located and migrating? | Order book snapshots, depth analysis | Daily updates |
| **Structural Trigger Analysis** | Breakout probability | Do conditions favor expansion or continued range? | Combined metrics across all categories | Pre-market daily |

| Analysis Type | Primary Focus | Key Questions | Data Requirements | Reporting Cadence |
|---|---|---|---|---|
| **Risk Environment** | Systemic vulnerability | What tail risks exist from leverage/liquidations? | Liquidation density maps, leverage data | Weekly deep-dive |
| **Narrative Filtering** | News relevance | Does news materially affect structure/funding? | Structural metrics before/after events | Event-driven |

*Note*: Regime identification uses 20-period lookback for accumulation/distribution patterns. Funding analysis tracks 7-day and 30-day rolling averages. Liquidity assessment monitors top 10% of order book by depth. Trigger analysis synthesizes ≥4 metrics for conviction ratings.

## Table XIV. Platform Operator Parameter Management Framework

| Parameter Type | Adjustment Objective | Trade-offs | Monitoring Metrics | Adjustment Triggers |
|---|---|---|---|---|
| **Margin Requirements** | Balance accessibility vs. systemic risk | Tighter = lower risk but compressed ranges | Liquidation frequency, cascade severity | Volatility regime changes |
| **Liquidation Algorithms** | Optimize deleveraging vs. market impact | Gradual = less panic; Aggressive = faster clearing | Slippage during liquidations, cascade duration | Extreme volatility events |
| **Funding Frequency** | Align derivative-spot parity | More frequent = tighter tracking; Less = lower friction | Basis persistence, arbitrage profitability | Persistent basis deviations |
| **Leverage Limits** | Constrain maximum position risk | Lower = safer; Higher = more attractive to traders | Average leverage utilization, user complaints | Risk events, regulatory changes |
| **Tick Size** | Balance precision vs. fragmentation | Smaller = more precise; Larger = better liquidity | Spread distribution, quote density | Liquidity regime shifts |
| **Insurance Fund Policy** | Protect against socialized losses | Larger = safer; Costs opportunity for users | Fund adequacy ratio, drawdown frequency | Major liquidation events |

*Note*: Parameter adjustments should be announced ≥24 hours in advance except during emergency risk events. Dynamic margin scaling can range from 20x (high volatility) to 100x (low volatility) based on 30-day realized volatility percentiles. Liquidation algorithm switches should occur at realized volatility exceeding 80th percentile of 90-day distribution.